\documentclass[a4paper]{jpconf}
\usepackage[utf8]{inputenc}
\usepackage{amsmath}
\usepackage{mathrsfs}
\usepackage[nopatch]{microtype}
\usepackage{booktabs}
\usepackage{url}
\usepackage{graphicx}
\usepackage[table]{xcolor}
\usepackage{colortbl}
\usepackage{multirow}
\usepackage{multicol}
\usepackage{verbatim}
\usepackage{color}
\usepackage{enumerate}
\usepackage{lscape}
\usepackage{gensymb}
\usepackage{amssymb}
\usepackage{diagbox}
\usepackage{array}
\usepackage{psfrag}
\usepackage[bottom]{footmisc} 
\usepackage[colorlinks,citecolor=blue,urlcolor=blue,bookmarks=false,hypertexnames=true]{hyperref}
\usepackage{helvet}
\usepackage{titlesec}
\titleformat{\chapter}
  {\normalfont\fontsize{16}{19}\sffamily\bfseries}
  {\thechapter}
  {1em}
  {}
\titleformat{\section}
  {\normalfont\fontsize{12}{17}\sffamily\bfseries}
  {\thesection}
  {1em}
  {}
\titleformat{\subsection}
  {\normalfont\fontsize{12}{17}\sffamily\bfseries\slshape}
  {\thesubsection}
  {1em}
  {}

\usepackage{fancyhdr}
\begin{document}
\title{On the effect of gravitational repulsion on geodesic deviation in the Schwarzschild-de Sitter spacetime}
\author{\textbf{Rohit Ghosh}$^{1,\,*}$, \textbf{Biplab Raychaudhuri}$^2$, \textbf{Aditya S. Mondal}$^3$ \textbf{and} \textbf{Mahasweta Bhattacharya}$^4$}
\small
\address{Department of Physics, Visva-Bharati, Santiniketan - 731235, West Bengal, India\\\vspace{0.25cm}
\textbf{E-mail:} $^1$\textbf{\href{mailto:3333372301@visva-bharati.ac.in}{3333372301@visva-bharati.ac.in}},  $^2$\textbf{\href{mailto:biplabphy@visva-bharati.ac.in}{biplabphy@visva-bharati.ac.in}}, $^3$\textbf{\href{mailto:adityas.mondal@visva-bharati.ac.in}{adityas.mondal@visva-bharati.ac.in}} \textbf{and} $^4$\textbf{\href{mailto:03333372303@visva-bharati.ac.in}{03333372303@visva-bharati.ac.in}}\\
$^*$Author to whom any correspondence should be addressed.\\\vspace{0.25cm}
Received June 2026}

\begin{abstract}
Here we discuss the effect of gravitational repulsion on geodesic deviation in the Schwarzschild-de Sitter spacetime. In particular, the geodesic deviation shows different behaviour inside or outside a \emph{critical surface}, which is the divider between attractive and repulsive regions of gravity. This in turn makes it possible to locate the critical surface and regions of gravitational repulsion and attraction from the behaviour of geodesic deviation.
\end{abstract}

\section{Introduction\vspace{0.25cm}}
\label{sec1}
The problem of gravitational repulsion or Hilbert repulsion \cite{loinger_marisco} has been a topic of discussion for many years. Newtonian gravity states that the gravitational force exerted on a mass is attractive in nature. While Newton's theory is limited to particles travelling slower than light in weak gravitational fields, Einstein's general relativity is applicable to arbitrary spacetimes and velocities. As demonstrated by McGruder III \cite{mcgruder1982}, gravitational repulsion of test particles with certain relativistic Schwarzschild velocities by gravitating point masses is allowed, without violating the basic tenet of general relativity. Droste \cite{droste}, Hilbert \cite{hilbert}, and Bauer \cite{bauer} independently discovered repulsive gravity in the Schwarzschild spacetime. McGruder III  and VanDerMeer \cite{mcgruder_vandermeer} provided the English translation of Droste's doctoral thesis. The conditions under which gravitational repulsion occurs have been explained by Treder and Fritze \cite{treder_fritze}, and McGruder III \cite{mcgruder2017}. A detailed study of this phenomenon has also been done by Arifov \cite{arifov}. Gravitational repulsion can accelerate particles to the highest cosmic ray energies known, as shown by McGruder III \cite{mcgruder2017}. The gravitational acceleration of elementary particles is slightly charge-dependent \cite{mcgruder1978}. But the contribution of such terms can be easily ignored as they require the existence of a highly charged particle of negligible mass, which is impossible according to current Particle Physics.

The possibility of repulsive gravity in the Reissner-Nordström spacetime has been discussed by Ponce de Leon \cite{poncedeleon} and Célérier, Santos, and Satheeshkumar \cite{célérier_santos_satheeshkumar}. Krori and Barua \cite{krori_barua} demonstrated that neutral test particles can be repelled in the Kerr spacetime and determined the circumstances in which this can occur. They investigated gravitational repulsion as perceived by a distant observer. Krori, Sarmah, and Goswami \cite{krori_sarmah_goswami} investigated the possibility of this effect in the Einstein-zero-mass scalar theory. As explained by Gorkavyi and Vasilkov \cite{gorkavyi_vasilkov}, repulsive forces appear to have been important in the past, based on the observable expansion of the universe. The concepts of the cosmological constant ($\Lambda$) \cite{matos} and dark energy \cite{ma_wang,haghani,harvey} may be useful in understanding the physical origin of gravitational repulsion. Gravitational repulsion is also a widely discussed phenomenon for asymptotically non-flat spacetimes. Recently, Luongo and Quevedo \cite{luongo_quevedo} have demonstrated the repulsive nature of gravity in regular black holes, using eigenvalues of the Riemann curvature tensor and the sign of the Ricci scalar. It has been demonstrated by Giacchini, Netto, and Modesto \cite{giacchini} that there exist singular higher-derivative curvature invariants for a wide class of black hole metrics that are prominently regular. Polo and Singh \cite{polo_singh} have examined the possibility of this effect in spacetimes possessing frame-dragging effect, using the radial acceleration of test particles.

Observational evidence from various cosmological probes — such as high redshift Type Ia supernovae \cite{perlmutter_turner_white}, the cosmic microwave background radiation \cite{dodelson_knox}, and the large-scale structure of the universe \cite{springel_frenk_white} — strongly suggests that the universe is not only expanding, but doing so at an accelerating rate. This accelerated expansion \cite{peebles_ratra, carroll} is in agreement with the theoretical framework of de Sitter spacetimes, which essentially have a positive cosmological constant ($\Lambda$). Harari and Luost\'o \cite{harari_luosto} showed that for Barriola-Vilenkin \cite{barriola_vilenkin} type global monopole, the effective mass is negative, producing a repulsive gravitational effect — a property shared with other topological defects such as domain walls and global strings. This repulsive gravity is conceptually linked to the cosmological constant ($\Lambda$), which represents the vacuum energy density of spacetime. Einstein's $\Lambda$, reinterpreted in modern cosmology as one of the candidates for dark energy, accounts for the accelerated expansion of the universe. However, the theoretical origin of $\Lambda$ remains elusive. Raychaudhuri, Rahaman and Kalam \cite{raychaudhuri_rahaman_kalam} proposed that phase transitions in the early universe, during which topological defects like monopoles formed, could generate an effective vacuum energy that manifests as $\Lambda$. Thus, both the negative effective mass of global monopoles and the repulsive gravity of dark energy may stem from the same underlying field-theoretic mechanism: the vacuum expectation energy associated with symmetry breaking at the early universe. This connection suggests that the global monopoles provide a microphysical analogue for the repulsive effects of the cosmological constant. It is also very interesting to deal with Anti-de Sitter (AdS) spacetimes as they are linked to several important theories, like the AdS/Conformal Field Theory (CFT) correspondence \cite{aharony_et_al} and the Randall–Sundrum braneworld scenario \cite{randall_sundrum,paul_sengupta}.

In this present work, we study the effect of the phenomenon of gravitational repulsion on the behaviour of geodesic deviation \cite{carroll_book,hobson,bażański_jaranowski}. We discuss such issue in the Schwarzschild-de Sitter spacetime. Geodesic deviation gives a direct physical meaning of spacetime curvature in general relativity. Consider two nearby test particles that are in free fall but were initially at rest relative to each other. In flat spacetime they would remain at a fixed separation, but in a curved one they generally do not: their separation changes with time, and this change is not an artifact of coordinates or of choosing a non-inertial frame. The particles experience a genuine relative acceleration, reflecting the fact that free-fall worldlines are not mutually parallel in a curved geometry. This relative acceleration represents the differential influence of gravitation across spacetime, responsible for characteristic stretching in one direction and compression in another. Thus, geodesic deviation expresses gravity in its invariant form: curvature controls how neighbouring freely falling trajectories converge or diverge, providing an experimentally meaningful probe of the local spacetime geometry.

In this work, we suggest a method to locate the region of gravitational repulsion from the behaviour of geodesic deviation in the Schwarzschild-de Sitter (SdS) spacetime. After finding a \emph{critical surface} in the SdS spacetime in Section \eqref{sec2}, we study the phenomenon of gravitational repulsion from the perspective of geodesic deviation in the SdS spacetime in Section \eqref{sec3}. The paper ends with some conclusive remarks in Section \eqref{sec4}.

\section{Existence of a \emph{critical surface} in the SdS spacetime\vspace{0.25cm}}
\label{sec2}
In this section we show that a critical surface exists for the SdS spacetime. The Schwarzschild metric in presence of a cosmological constant $(\Lambda)$, in spherical polar coordinates (and in geometrized units: $c=G=1$) is given by:
\begin{equation}
\label{eqn1}
\text{d}s^{2}=-\,f\left(r\right)\text{d}t^{2}+f\left(r\right)^{-1}\text{d}r^{2}+r^{2}\text{d}\theta^{2}+r^{2}\sin^2\theta\text{d}\phi^2\text{.}
\end{equation}
Here, $f(r)=1-\frac{2M}{r}-\frac{\Lambda r^2}{3}$, $M$ is a mass parameter and $\Lambda$ is known as the cosmological constant, which makes the underlying spacetime asymptotically non-flat. $\Lambda=0$, $\Lambda>0$ and $\Lambda<0$ for the Schwarzschild, the SdS and the Schwarzschild–Anti-de Sitter (SAdS) spacetimes, respectively.\\\\
The Lagrangian function of a test particle associated with the spacetime in Eq.~\eqref{eqn1} is given by:
\begin{equation}
\label{eqn2}
\mathscr{L} = g_{\mu\nu}\frac{\text{d}x^{\mu}}{\text{d}\lambda}\frac{\text{d}x^{\nu}}{\text{d}\lambda}\\
 = -\,f(r)\left(\frac{\text{d}t}{\text{d}\lambda}\right)^2+f(r)^{-1}\left(\frac{\text{d}r}{\text{d}\lambda}\right)^2+r^2\left(\frac{\text{d}\theta}{\text{d}\lambda}\right)^2+r^2\sin^2\theta\left(\frac{\text{d}\phi}{\text{d}\lambda}\right)^2\text{.}
\end{equation}
where, $\lambda$ is an affine parameter. Using the spherical symmetry of the metric, we can consider the motion along the equatorial plane ($\theta=\frac{\pi}{2}$). As a consequence, the Lagrangian function reduces to:
\begin{equation}
\label{eqn3}
\mathscr{L} = -\,f(r)\left(\frac{\text{d}t}{\text{d}\lambda}\right)^{2}+f(r)^{-1}\left(\frac{\text{d}r}{\text{d}\lambda}\right)^{2}+r^{2}\left(\frac{\text{d}\phi}{\text{d}\lambda}\right)^{2}\text{.}
\end{equation}
 The fact that $\mathscr{L}$ is explicitly independent of $t$ and $\phi$ ensures the conservation of $p_t$ and $p_\phi$ (i.e., the `t' and `$\phi$' components of the four momentum) of the test particle along the geodesics. This leads to the first integrals of the temporal and the azimuthal equations, given by:
\begin{equation}
\label{eqn4}
p_t = -\,f(r)\frac{\text{d}t}{\text{d}\lambda}= -\,k\,\,\text{(const.).}
\end{equation}

\begin{equation}
\label{eqn5}
\text{and\quad} p_\phi = r^2\frac{\text{d}\phi}{\text{d}\lambda}= h\,\,\text{(const.).}
\end{equation}
Here, we have used $p^{\mu}=\frac{\text{d}x^{\mu}}{\text{d}\lambda}\Rightarrow
p_{\mu}=g_{\mu\nu}p^{\nu}=g_{\mu\nu}\frac{\text{d}x^{\nu}}{\text{d}\lambda}$.\\
\\
Here, $k$ and $h$ are the energy and the angular momentum per unit mass of the test particle, respectively. Eqs.~\eqref{eqn4} and \eqref{eqn5} immediately lead to:
\begin{equation}
\label{eqn6}
\frac{\text{d}t}{\text{d}\lambda}=kf(r)^{-1}
\end{equation}
and
\begin{equation}
\label{eqn7}
\frac{\text{d}\phi}{\text{d}\lambda}=\frac{h}{r^2}\text{,}
\end{equation}
respectively. Moreover, from the invariant length of the 4-momentum, we have:
$g^{\mu\nu}p_{\mu}p_{\nu}=\epsilon$, which owing to the previous expressions, ultimately boils down to:\\
\begin{equation}
\label{eqn8}
\frac{\text{d}r}{\text{d}\lambda}=\left\{ k^{2}+\left(\epsilon-\frac{h^{2}}{r^{2}}\right)f(r)\right\} ^{\frac{1}{2}}\text{.}
\end{equation}
\\
Here, $\epsilon=-1, 0$ and $1$ for time-like particles (tardyons), light-like particles (photons) and space-like particles (tachyons), respectively. In the present work, we consider only the time-like or the material particles for which $\epsilon =-1$. We consider the motion of such particles with respect to a local observer with proper time given by $\tau$. The radial coordinate $r=r(\tau)$ in this case is defined as the position of the test particle according to the local observer.\\\\
We now parametrize the trajectory by the proper time. Hence, $\lambda=\tau$ and Eq.~\eqref{eqn8} reduces to:
\begin{equation}
\label{eqn9}
\frac{\text{d}r}{\text{d}\tau}=\left\{k^{2}+\left(\epsilon-\frac{h^{2}}{r^{2}}\right)f(r)\right\}^{\frac{1}{2}}\text{,}
\end{equation}
which represents the radial speed of the test particle as measured by the local observer.\\\\
For tardyons ($\epsilon=-1$) in radial motion (i.e. $h=0$), Eq.~\eref{eqn9} reduces to:
\begin{equation}
\label{eqn10}
\frac{\text{d}r}{\text{d}\tau}=\left\{k^2-f(r)\right\}^{\frac{1}{2}}\text{.}
\end{equation}

In order to determine the nature of gravity, we need to check the sign of the radial acceleration of the test particle; in particular, positive and negative signs represent gravitational repulsion and attraction, respectively. For this purpose, we first differentiate Eq.~\eqref{eqn9} with respect to $\tau$ to get:\\
\begin{equation}
\label{eqn11}
\frac{\text{d}^2r}{\text{d}\tau^2}=\epsilon\left(\frac{M}{r^2}-\frac{\Lambda r}{3}\right)+\frac{h^2}{r^3}\left(1-\frac{3M}{r}\right)\text{.}
\end{equation}\\
For such parametrization, Eq.~\eqref{eqn7} gives:
\begin{equation}
\label{eqn12}
\frac{\text{d}\phi}{\text{d}\tau}=\frac{h}{r^2}\text{.}
\end{equation}
Using Eqs.~\eqref{eqn11} and \eqref{eqn12}, the radial acceleration of the test particle as measured by the local observer is given by:
\begin{equation}
\label{eqn13}
a=\frac{\text{d}^2r}{\text{d}\tau^2}-r\left(\frac{\text{d}\phi}{\text{d}\tau}\right)^2=\epsilon\left(\frac{M}{r^2}-\frac{\Lambda r}{3}\right)-\frac{3Mh^2}{r^4}\text{.}
\end{equation}\\
For tardyons ($\epsilon=-1$) in radial motion ($h=0$)\footnote[2]{For transverse motion ($h\ne0$ and $\frac{\text{d}r}{\text{d}\tau}=0$), $a=-r\left(\frac{\text{d}\phi}{\text{d}\tau}\right)^2=-\frac{h^2}{r^3}$, which is generally negative and also independent of $\Lambda$. Hence, the test particles only experience gravitational attraction.}, Eq.~\eqref{eqn13} gives:
\begin{equation}
\label{eqn14}
a=-\,\frac{M}{r^2}\left(1-\frac{\Lambda r^3}{3M}\right)\text{.}
\end{equation}
In this regard, three different cases need to be mentioned depending on the sign of $\Lambda$ ---

\begin{itemize}
\item \underline{\textbf{For} $\boldsymbol{\Lambda=0}$} \textbf{:} Here Eq.~\eqref{eqn14} reduces to:
\begin{equation}
\label{eqn15}
a_\text{Sch.}=-\,\frac{M}{r^2}\text{.}
\end{equation}
Hence in the Schwarzschild spacetime, for radial motion, test particles (tardyons) only experience gravitational attraction according to the local observer. Eq.~\eqref{eqn15} is in agreement with what McGruder III \cite{mcgruder1982} had implied.
\item \underline{\textbf{For} $\boldsymbol{\Lambda>0}$} \textbf{:}
According to Eq.~\eqref{eqn14}, here, i.e., in the SdS spacetime, for radial motion, test particles experience gravitational repulsion and gravitational attraction for $r>\sqrt[3]{3M/\Lambda}$ and $r<\sqrt[3]{3M/\Lambda}$, respectively according to the local observer. $a=0$ for $r=\sqrt[3]{3M/\Lambda}$ (consequently, the radial speed is minimum for $r=\sqrt[3]{3M/\Lambda}$). The repulsion is solely due to $\Lambda$. Therefore, from the perspective of repulsion and attraction, we may consider $r=\sqrt[3]{3M/\Lambda}$ as a critical surface set by the spacetime.
\item \underline{\textbf{For} $\boldsymbol{\Lambda<0}$} \textbf{:} According to Eq.~\eqref{eqn14}, for this case we see that the product of both the terms ultimately leads to a negative sign. Thus we may argue that in the SAdS spacetime, test particles, moving along the radial direction, only experience gravitational attraction according to the local observer.
\end{itemize}

Based on the above discussions, we present Table (\ref{tab1}) demonstrating a comparative study of the nature of gravity acting on test particles as perceived by a local observer in Schwarzschild-like spacetimes [as defined by Eq.~\eqref{eqn1}].\\

\begin{table}[!htb]
\centering
\caption{\textbf{Nature of gravity experienced by radially moving test particles in Schwarzschild-like spacetimes as perceived by a local observer\vspace{0.25cm}}}
\label{tab1}
\begin{tabular}{|c|c|c|}
\hline
\rowcolor{blue!30}
  \begin{tabular}[c]{@{}l@{}}\boldmath{$\Lambda=0$}\end{tabular} &
  \begin{tabular}[c]{@{}l@{}}\boldmath{$\Lambda>0$}\end{tabular} &
  \begin{tabular}[c]{@{}l@{}}\boldmath{$\Lambda<0$}\end{tabular}\\ \hline
\rowcolor{blue!5!white}
  Attractive &
  \begin{tabular}[c]{@{}l@{}}\textbf{(i)} Repulsive for $r>\sqrt[3]{\frac{3M}{\Lambda}}$\\\textbf{(ii)} Attractive for $r<\sqrt[3]{\frac{3M}{\Lambda}}$\\ \textbf{(iii)} $a=0$ for $r=\sqrt[3]{\frac{3M}{\Lambda}}$\end{tabular} &
  Attractive \\\hline
\rowcolor{blue!5!white}
\end{tabular}
\end{table}
It is evident from Table (\ref{tab1}) that radially moving test particles experience gravitational repulsion only when they are in the SdS spacetime. This repulsive behaviour (with respect to the local observer) is solely due the presence of a positive cosmological constant ($\Lambda$).\\

As shown by Hilbert \cite{hilbert}, gravitational repulsion in the pure Schwarzschild spacetime originates from considering $t$, instead of $\tau$ as our time-like coordinate. Due to the asymptotic flatness of this spacetime, it is physically possible to associate a distant observer with proper time $t$. But this same treatment cannot be done for the SdS or the SAdS spacetimes, as both of them are asymptotically non-flat due to the presence of non-zero cosmological constant $\Lambda$. So for further analysis, we will only consider the time-like coordinate $\tau$. As we have already seen that radially moving test particles experience gravitational repulsion only in the SdS spacetime (with respect to $\tau$), in the next section, we will study the effect of repulsive gravity on the acceleration of the geodesic deviation 4-vectors in the SdS spacetime.

\section{Implications of gravitational repulsion on geodesic deviation in the SdS spacetime\vspace{0.25cm}}
\label{sec3}
We consider two neighbouring geodesics (two neighbouring particles having the same energy per unit mass, $k$), affinely parametrized by $\lambda$ in the SdS spacetime. Let $\xi^\mu(\lambda)$ be the 4-vector connecting the two geodesics. The equations of geodesic deviation are then given by:
\begin{equation}
\label{eqn16}
\frac{\text{D}^2\xi^\mu}{\text{D}\lambda^2}+R^\mu_{\rho\nu\sigma}\xi^\nu\frac{\text{d}x^\rho}{\text{d}\lambda}\frac{\text{d}x^\sigma}{\text{d}\lambda}=0\text{.}
\end{equation}
where, $\frac{\text{D}}{\text{D}\lambda}$ denotes covariant derivative.\\\\
For $\mu=0$ in Eq.~\eqref{eqn16} we have the $t$-equation:
\begin{equation}
\label{eqn17}
\frac{\text{D}^2\xi^t}{\text{D}\lambda^2}+R^t_{rtr}\frac{\text{d}r}{\text{d}\lambda}\left(\frac{\text{d}r}{\text{d}\lambda}\xi^t-\frac{\text{d}t}{\text{d}\lambda}\xi^r\right)+R^t_{\theta t\theta}\frac{\text{d}\theta}{\text{d}\lambda}\left(\frac{\text{d}\theta}{\text{d}\lambda}\xi^t-\frac{\text{d}t}{\text{d}\lambda}\xi^\theta\right)
+R^t_{\phi t\phi}\frac{\text{d}\phi}{\text{d}\lambda}\left(\frac{\text{d}\phi}{\text{d}\lambda}\xi^t-\frac{\text{d}t}{\text{d}\lambda}\xi^\phi\right)=0\text{.}
\end{equation}
For $\mu=1$ in Eq.~\eqref{eqn16} we have the $r$-equation:
\begin{equation}
\label{eqn18}
\frac{\text{D}^2\xi^r}{\text{D}\lambda^2}+R^r_{trt}\frac{\text{d}t}{\text{d}\lambda}\left(\frac{\text{d}t}{\text{d}\lambda}\xi^r-\frac{\text{d}r}{\text{d}\lambda}\xi^t\right)+R^r_{\theta r\theta}\frac{\text{d}\theta}{\text{d}\lambda}\left(\frac{\text{d}\theta}{\text{d}\lambda}\xi^r-\frac{\text{d}r}{\text{d}\lambda}\xi^\theta\right)
+R^r_{\phi r\phi}\frac{\text{d}\phi}{\text{d}\lambda}\left(\frac{\text{d}\phi}{\text{d}\lambda}\xi^r-\frac{\text{d}r}{\text{d}\lambda}\xi^\phi\right)=0\text{.}
\end{equation}
For the equatorial plane ($\theta=\frac{\pi}{2}$) and radial motion ($\frac{\text{d}\phi}{\text{d}\lambda}=0$), Eqs.~\eqref{eqn17} and \eqref{eqn18} reduce to:
\begin{equation}
\label{eqn19}
\frac{\text{D}^2\xi^t}{\text{D}\lambda^2}=-R^t_{rtr}\frac{\text{d}r}{\text{d}\lambda}\left(\frac{\text{d}r}{\text{d}\lambda}\xi^t-\frac{\text{d}t}{\text{d}\lambda}\xi^r\right)
\end{equation}
and
\begin{equation}
\label{eqn20}
\frac{\text{D}^2\xi^r}{\text{D}\lambda^2}=-R^r_{trt}\frac{\text{d}t}{\text{d}\lambda}\left(\frac{\text{d}t}{\text{d}\lambda}\xi^r-\frac{\text{d}r}{\text{d}\lambda}\xi^t\right)\text{,}
\end{equation}
respectively. Using
\begin{equation}
\label{eqn21}
\frac{\text{D}^2\xi^\mu}{\text{D}\lambda^2}=\frac{\text{d}^2\xi^\mu}{\text{d}\lambda^2}+(\partial_\sigma\Gamma^\mu_{\nu\rho})\xi^\nu\frac{\text{d}x^\rho}{\text{d}\lambda}\frac{\text{d}x^\sigma}{\text{d}\lambda},
\end{equation}
Eqs.~\eqref{eqn20} and \eqref{eqn21} can be rewritten as:
\begin{equation}
\label{eqn22}
\frac{\text{d}^2\xi^t}{\text{d}\lambda^2}=-(R^t_{rtr}+\partial_r\Gamma^t_{tr})\left(\frac{\text{d}r}{\text{d}\lambda}\right)^2\xi^t+(R^t_{rtr}-\partial_r\Gamma^t_{tr})\frac{\text{d}t}{\text{d}\lambda}\frac{\text{d}r}{\text{d}\lambda}\xi^r
\end{equation}
and
\begin{equation}
\label{eqn23}
\frac{\text{d}^2\xi^r}{\text{d}\lambda^2}=(R^r_{trt}-\partial_r\Gamma^r_{tt})\frac{\text{d}t}{\text{d}\lambda}\frac{\text{d}r}{\text{d}\lambda}\xi^t-\left\{R^r_{trt}\left(\frac{\text{d}t}{\text{d}\lambda}\right)^2+(\partial_r\Gamma^r_{rr})\left(\frac{\text{d}r}{\text{d}\lambda}\right)^2\right\}\xi^r,
\end{equation}
respectively. Here, $\Gamma^\mu_{\nu\alpha}$ are the Christoffel symbols of the $2^{\text{nd}}$ kind and $R^\mu_{\nu\alpha\beta}$ is the Riemann tensor.\\\\
At this point, we re-parametrize the system using proper time by setting $\lambda=\tau$. For the metric given by Eq.~\eqref{eqn1}, Eqs.~\eqref{eqn22} and \eqref{eqn23} (after re-parametrization) can be expressed as:
\begin{equation}
\label{eqn24}
\frac{\text{d}^2\xi^t}{\text{d}\tau^2}=C_{00}(r)\xi^t+C_{01}(r)\xi^r
\end{equation}
and
\begin{equation}
\label{eqn25}
\frac{\text{d}^2\xi^r}{\text{d}\tau^2}=C_{10}(r)\xi^t+C_{11}(r)\xi^r\text{,}
\end{equation}
respectively. Here,
\begin{equation}
\label{eqn26}
C_{00}(r)=\frac{2}{9}\left[\left\{\frac{\Lambda r^3-3M}{r^2f(r)}\right\}\left\{k^{2}-f(r)\right\}^{\frac{1}{2}}\right]^2\geqslant0\text{,}
\end{equation}
\begin{equation}
\label{eqn27}
C_{01}(r)=\frac{2k\left\{k^{2}-f(r)\right\}^{\frac{1}{2}}(\Lambda r^2-6\Lambda Mr+6M/r-9M^2/r^2)}{3r^2\{f(r)\}^3}\text{,}
\end{equation}
\begin{equation}
\label{eqn28}
C_{10}(r)=-\frac{f(r)}{\left\{1-f(r)/k^2\right\}^\frac{1}{2}}C_{00}(r)\leqslant0
\end{equation}
and
\begin{equation}
\begin{split}
\label{eqn29}
C_{11}(r)=\frac{1}{\{f(r)\}^2}\left\{\frac{4M^3}{r^5}-\frac{2M^2(k^2+3)}{r^4}+\frac{2M}{r^3}+\frac{6\Lambda M^2}{r^2}+\frac{4\Lambda M(k^2/3-1)}{r}\right.&+\frac{\Lambda}{3}+\frac{2\Lambda^2Mr}{3}\\
&\left.-\frac{2\Lambda^2k^2r^2}{9}-\frac{\Lambda^3r^4}{27}\right\}\text{.}
\end{split}
\end{equation}
In Eqs.~\eqref{eqn26}-\eqref{eqn29}, the 1st and the 2nd subscripts represent the component type of the acceleration of the geodesic deviation 4-vector and the component type of the geodesic deviation 4-vector along which it is aligned, respectively.

In section ({\ref{sec2}}), we have seen that in the SdS spacetime there exists a critical surface for $a=0$ ($r=\sqrt[3]{3M/\Lambda}$), which acts as an interface of gravitational attraction and repulsion for radially moving test particles with respect to a local observer. We will now examine the behaviour of the acceleration of the geodesic deviation 4-vector [Eqs.~\eqref{eqn24} and \eqref{eqn25}] on and away from the critical surface.\\\\
From Eqs.~\eqref{eqn26}-\eqref{eqn29}, it follows that:
\begin{equation*}
C_{00}\,\left(\sqrt[3]{3M/\Lambda}\right)=C_{10}\,\left(\sqrt[3]{3M/\Lambda}\right)=0,
\end{equation*}
\begin{equation*}
C_{01}\,\left(\sqrt[3]{3M/\Lambda}\right)=\frac{2\Lambda k\left\{k^2-f\left(\sqrt[3]{3M/\Lambda}\right)\right\}^\frac{1}{2}}{\left\{f\left(\sqrt[3]{3M/\Lambda}\right)\right\}^2}\quad\text{and}
\end{equation*}
\begin{equation*}
C_{11}\,\left(\sqrt[3]{3M/\Lambda}\right)=\Lambda.
\end{equation*}
Using the last four relations at $r=\sqrt[3]{3M/\Lambda}$, Eqs.~\eqref{eqn24} and \eqref{eqn25}, reduce to:
\begin{equation}
\label{eqn30}
\left.\frac{\text{d}^{2}\xi^t}{\text{d}\tau^2}\right|_{r=\sqrt[3]{3M/\Lambda}}=\frac{2\Lambda k \left\{k^2-f\left(\sqrt[3]{3M/\Lambda}\right)\right\}^\frac{1}{2}}{\left\{f\left(\sqrt[3]{3M/\Lambda}\right)\right\}^2}\left.\xi^r\right|_{r=\sqrt[3]{3M/\Lambda}}
\end{equation}
and
\begin{equation}
\label{eqn31}
\left.\frac{\text{d}^{2}\xi^r}{\text{d}\tau^2}\right|_{r=\sqrt[3]{3M/\Lambda}}=\Lambda\left.\xi^r\right|_{r=\sqrt[3]{3M/\Lambda}}\text{,}
\end{equation}
respectively.\\\\
The above two equations clearly indicate that both the components of the acceleration of the geodesic deviation 4-vector on the critical surface ($r=\sqrt[3]{3M/\Lambda}$), lie toward $\xi^r$. Since $C_{00}(r)$ and $C_{10}(r)$ show a distinct behaviour on the critical surface, we will now analyze them graphically to understand the nature of gravity.
\begin{figure}[ht]
\centering
\includegraphics[scale=0.4]{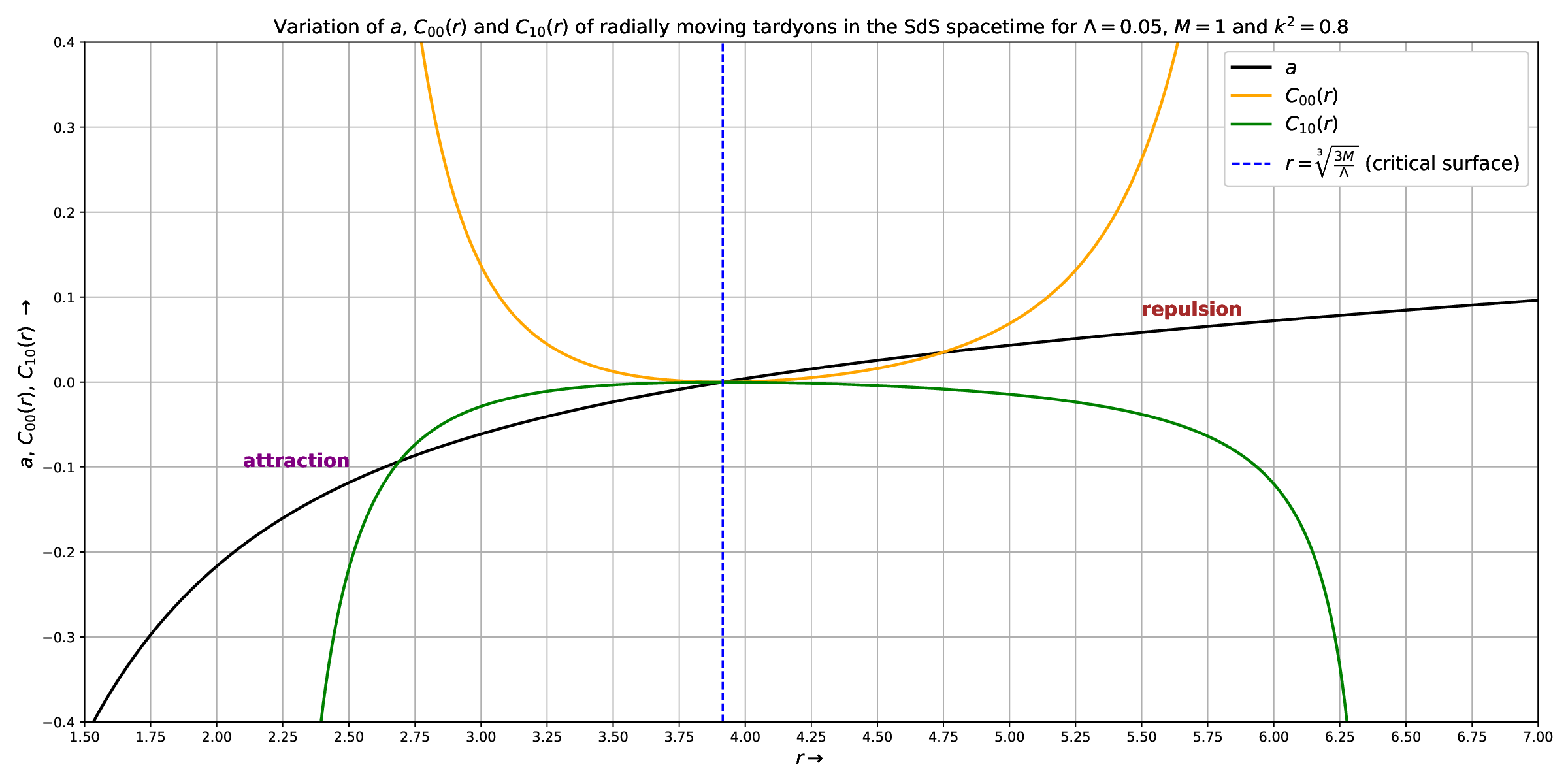}
\caption{\textbf{Variation of} \boldmath{$a$} \textbf{[Eq.~\eqref{eqn14}],} \boldmath{$C_{00}(r)$} \textbf{[Eq.~\eqref{eqn26}] and} \boldmath{$C_{10}(r)$} \textbf{[Eq.~\eqref{eqn28}] of radially moving test particles (tardyons) in the SdS spacetime for} \boldmath{$\Lambda=0.05$}, \boldmath{$M=1$} \textbf{and} \boldmath{$k^2=0.8$}}
\label{fig1}
\end{figure}\\
The characteristic variation of $a$, $C_{00}(r)$ and $C_{10}(r)$ are plotted in Figure~\eqref{fig1} for some representative values of $\Lambda$, $M$ and $k^2$. This demonstrates that $C_{00}(r)$ has a minimum and $C_{10}(r)$ has a maximum at $r=\sqrt[3]{3M/\Lambda}$, respectively. The value of each of them is also zero there. The attractive and repulsive gravitational regions lie inside and outside of the critical surface, respectively. This is in agreement with our previous arguments.\\\\
 Table (\ref{tab2}) shows the behaviour of the three functions on and away from the critical surface and depicts a possible method to locate the critical surface and the attractive/repulsive regions in the SdS spacetime.
\begin{table}[ht]
\centering
\caption{\textbf{Behaviour of }\boldmath{$a$} \textbf{[Eq.~\eqref{eqn14}],} 
\boldmath{$C_{00}(r)$} \textbf{[Eq.~\eqref{eqn26}] and} 
\boldmath{$C_{10}(r)$} \textbf{[Eq.~\eqref{eqn28}] of radially moving test particles in the SdS spacetime}\vspace{0.25cm}}
\label{tab2}
\begin{tabular}{|c|c|c|c|}
\hline
\rowcolor{blue!30}
\diagbox[width=7em]{\textbf{Function}}{\textbf{Region}} &
  \boldmath{$r<\sqrt[3]{\frac{3M}{\Lambda}}$} &
  \boldmath{$r=\sqrt[3]{\frac{3M}{\Lambda}}$} &
  \boldmath{$r>\sqrt[3]{\frac{3M}{\Lambda}}$} \\ 
\hline
\cellcolor{blue!30}\boldmath{$a$} & \cellcolor{blue!5!white}-ve and increasing with $r$ & \cellcolor{blue!5!white}Zero &\cellcolor{blue!5!white}+ve and increasing with $r$\\
\cline{1-4}
\cellcolor{blue!30}\boldmath{$C_{00}(r)$} & \cellcolor{blue!5!white}+ve and decreasing with $r$ & \cellcolor{blue!5!white}Zero (minimum) &\cellcolor{blue!5!white}+ve and increasing with $r$\\
\cline{1-4}
\cellcolor{blue!30}\boldmath{$C_{10}(r)$} & \cellcolor{blue!5!white}-ve and increasing with $r$ & \cellcolor{blue!5!white}Zero (maximum) & \cellcolor{blue!5!white}-ve and decreasing with $r$\\
\hline
\end{tabular}
\end{table}

\section{Conclusions\vspace{0.25cm}}
\label{sec4}
We have examined the conditions of possible gravitational repulsion with respect to local observers in the Schwarzschild spacetime in the presence of cosmological constant ($\Lambda$). It turns out that for a local observer (with proper time $\tau$) the repulsive nature of gravity may occur only for $\Lambda>0$, depending on the test particle's position. Here, it deserves mentioning that the Schwarzschild-de Sitter spacetime manifests the repulsive/attractive nature through a \emph{critical surface} of radius $\sqrt[3]{3M/\Lambda}$. In particular, the test particle \emph{beyond} the critical surface feels a \emph{repulsive} kind of gravity and an \emph{attractive} kind of gravity \emph{inside} the critical surface. Moreover, for $\Lambda\leqslant0$, gravity shows an attractive nature with respect to a local observer. On the critical surface, the components of the acceleration of the geodesic deviation 4-vector become parallel to the radial direction $\xi^r$ and this can be used to identify the critical surface and the nature of gravity. Thus, the present work establishes that gravitational repulsion in the Schwarzschild–de Sitter spacetime leaves a distinct signature not only on the geodesic motion of an individual radially outgoing test particle, but also on the relative acceleration of neighbouring radially outgoing geodesics.

\section*{Acknowledgement\vspace{0.25cm}}
RG would like to thank Dr. Tanmoy Paul, Department of Physics, Visva-Bharati, Santiniketan - 731235, West Bengal, India, for his valuable suggestions regarding the interpretation of the results.

\section*{References\vspace{0.25cm}}

\end{document}